\documentclass[twocolumn,floatfix,eqsecnum,aps,prb]{revtex4-2}
\usepackage{graphicx}
\usepackage{amsmath}
\usepackage{amssymb}
\usepackage{bm}
\usepackage{hyperref}

\usepackage{color}

\begin{document}

\title{Chiral edge mode for single-cone Dirac fermions}
\author{C. W. J. Beenakker}
\affiliation{Instituut-Lorentz, Universiteit Leiden, P.O. Box 9506, 2300 RA Leiden, The Netherlands}
\date{August 2024}

\begin{abstract}
We study the appearance of a chiral edge mode on the two-dimensional (2D) surface of a 3D topological insulator (TI). The edge mode appears along the 1D boundary with a magnetic insulator (MI), dependent on the angle $\theta$ which the magnetization $M$ makes with the normal to the surface and on the chemical potential mismatch $\delta\mu$ across the TI--MI interface (assuming $1>\delta\mu/M\equiv\sin\phi)$. The propagation along the interface is chiral, with velocity $v\cos(\theta-\phi)$ smaller than the Dirac fermion velocity $v$. In momentum space the edge mode is an arc state, extending over the finite momentum interval that connects the Dirac point of the gapless Dirac fermions with the magnetic band gap. An electric field parallel to the boundary pumps charge between TI and MI via this arc state.
\end{abstract}
\maketitle

\section{Introduction}

\begin{figure}[tb]
\centerline{\includegraphics[width=0.9\linewidth]{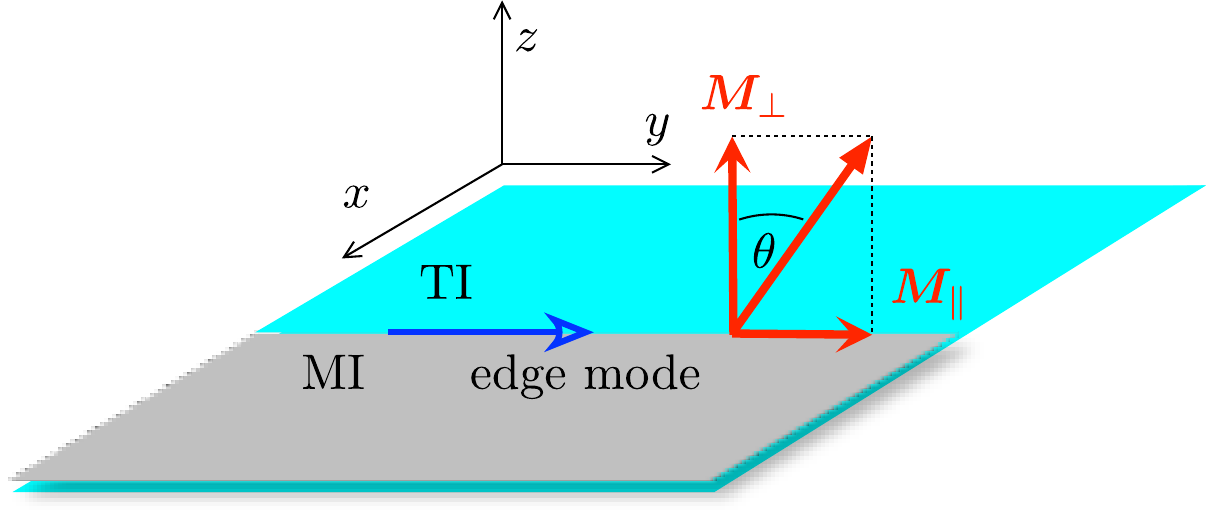}}
\caption{Top view of the 2D surface of a 3D topological insulator (TI), partially covered by a magnetic insulator (MI). A chiral edge mode propagates along the TI--MI boundary if the magnetization $\bm{M}$ has an in-plane component $\bm{M}_\parallel$ parallel to the boundary. The sum $\bm{M}_\perp+\bm{M}_\parallel$ of perpendicular and parallel components makes an angle $\theta$ with the normal to the surface.
}
\label{fig_diagram}
\end{figure}

The massless Dirac fermions on the two-dimensional (2D) surface of a topological insulator (TI) cannot be confined by electrostatic potentials, one needs to break time-reversal symmetry to remove the topological protection of the gapless Dirac cone \cite{Has10,Qi11}. For that purpose one can deposit a magnetic insulator (MI) on the TI, for example EuS on Bi$_2$Se$_3$ \cite{Kat16,Wan23}, which confines the Dirac fermions via the magnetic exchange interaction $\bm{M}\cdot\bm{\sigma}$. 

Such a heterostructure provides a condensed matter realization of the ``neutrino billiard'' studied in 1987 by Berry and Mondragon \cite{Ber87}, without the ``fermion-doubling'' complication from intervalley scattering that is present in graphene \cite{Pon08}. Even earlier, in the particle-physics context, the infinite-mass limit $M\rightarrow\infty$ was used as a ``bag model'' to confine relativistic fermions \cite{Cho74,Joh75}.

A magnetization $M$ perpendicular to the TI surface [in-plane momentum $\bm{p}=(p_x,p_y)$] adds a mass term $M\sigma_z$ to the 2D Dirac Hamiltonian $
H=v\bm{p}\cdot\bm{\sigma}$, thereby confining the low-energy excitations to regions where $M=0$. The mass boundary reflects the massless electrons without binding them to the interface: a bound state appears along an interface where $M$ changes sign \cite{Mar08}, but the interface with an $M=0$ region does not support an edge mode.

\begin{figure}[tb]
\centerline{\includegraphics[width=0.8\linewidth]{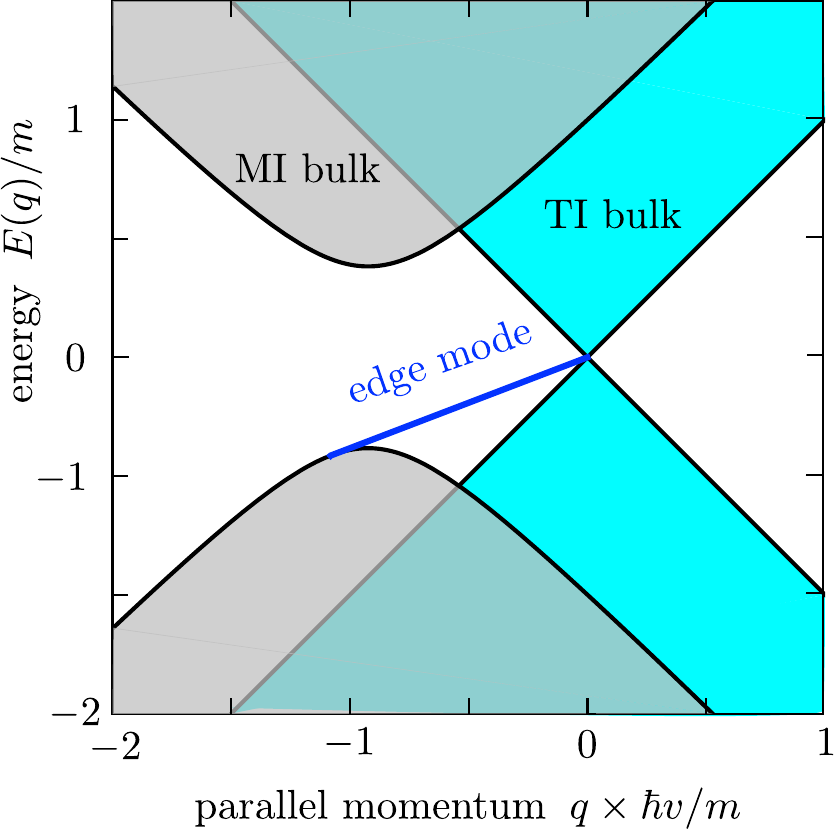}}
\caption{Band structure of the TI--MI surface, for an angle $\theta=3\pi/8$ of the magnetization vector $(0,m\sin\theta,m\cos\theta)$ with the normal to the surface. The edge mode with dispersion \eqref{edgemode} connects the bulk bands on the two sides of the interface. In this plot there is no chemical  potential mismatch between TI and MI ($\delta\mu=0$).
}
\label{fig_plumbing}
\end{figure}

Here we show that a chiral edge mode will appear if the magnetization is tilted away from the normal, so that it has both a normal component $M_\perp$ and an in-plane component $M_\parallel$ parallel to the interface (see Fig.\ \ref{fig_diagram}). The edge mode is an ``arc state'', ranging over a finite momentum interval (see Fig.\ \ref{fig_plumbing}), as a 2D analogue of the surface Fermi arc familiar from 3D Weyl semimetals \cite{Tur13,Yan17}. The edge mode at the TI-MI interface also has an analogue in graphene nanoribbons \cite{Akh08} --- but it is chiral rather than helical because of the absence of fermion doubling. The edge velocity $v_{\rm edge}=v\cos\theta$ vanishes if the magnetization is fully aligned with the interface. The TI edge mode is then the analogue of the dispersonless zigzag edge mode in graphene.  

Without the $M_\parallel$ component a chiral edge mode can be induced by a chemical potential mismatch $\delta\mu=\mu_{\rm MI}-\mu_{\rm TI}$. For $|\delta\mu|<|M|$ the angle $\phi=\arcsin(\delta\mu/M)$ determines the edge state velocity $v_{\rm edge}=v\cos(\theta-\phi)$, so it has the same effect as a rotation of the magnetization. 

\section{Chiral edge mode at a TI--MI boundary}

\subsection{Geometry}

The surface electrons in the geometry of Fig.\ \ref{fig_diagram} are described by the 2D Dirac Hamiltonian 
\begin{equation}
H=vp_x\sigma_x+vp_y\sigma_y+\bm{M}\cdot\bm{\sigma},\;\;\bm{p}=-i\hbar\partial/\partial\bm{r}.
\label{Hdef}
\end{equation}
The magnetization vector is $\bm{M}=(M_x,M_y,M_z)$, the Pauli spin matrices are $\bm{\sigma}=(\sigma_x,\sigma_y,\sigma_z)$, and $v$ is the Fermi velocity. We consider a boundary along the $y$-axis, with $\bm{M}(x)=0$ for $x<0$. We take the same chemical potential in TI and MI, the effect of a chemical potential mismatch is considered in Sec.\ \ref{sec_deltamu}.

We assume translational invariance along the $y$-axis, $\bm{M}=\bm{M}(x)$. The $x$-component $M_x$ of the magnetization (along $\bm{n}$) can be gauged away, it plays no role. The relevant components are $M_z=m\cos\theta\equiv M_\perp$ (perpendicular to the surface) and $M_y=m\sin\theta\equiv M_\parallel$ (in-plane parallel to the boundary). We take $m>0$ constant for $x>0$. For ease of notation we set $\hbar$ and $v$ equal to unity in most equations.

The momentum $p_y=q$ along the boundary is conserved. The eigenvalue problem for given $q$ is
\begin{equation}
-i\sigma_x\psi'(x)=\begin{cases}
(E-q\sigma_y)\psi(x)&x<0,\\
\bigl(E-(q+m\sin\theta)\sigma_y\\
-m\sigma_z\cos\theta\bigr)\psi(x),&x>0.
\end{cases}\label{dpsidx1}
\end{equation}
The solution is of the form
\begin{equation}
\psi(x)=\begin{cases}
e^{\Xi_0 x}\psi(0),&x<0,\\
e^{(\Xi_0+\Xi_{\rm M})x}\psi(0),&x>0,
\end{cases}
\end{equation}
with matrices
\begin{equation}
\begin{split}
&\Xi_0=iE\sigma_x+q\sigma_z,\\
&\Xi_{\rm M}=m(\sigma_z\sin\theta-\sigma_y\cos\theta).
\end{split}\label{Xi0XiMdef}
\end{equation}

\subsection{Edge mode dispersion}
\label{sec_edgemodedispersion}

We seek a solution that decays both for $x\rightarrow\infty$ and for $x\rightarrow-\infty$. The matrix $\Xi_0$ has eigenvalues
\begin{equation}
\pm\xi_0=\pm\sqrt{q^2-E^2},
\end{equation}
with \textit{left} eigenvectors
\begin{equation}
v_\pm=(q\pm\xi_0,-iE).
\end{equation}
The matrix $\Xi_0+\Xi_{\rm M}$ has eigenvalues 
\begin{equation}
\xi_{\rm M}=\pm\sqrt{q^2-E^2+m^2+2mq\sin\theta}
\end{equation}
with \textit{right} eigenvectors
\begin{equation}
u_\pm=q+m\sin\theta\pm\xi_{\rm M},iE-im\cos\theta).
\end{equation}

For a two-sided decay we require that $\xi_0$ and $\xi_{\rm M}$ are both real and $v_-$ is orthogonal to $u_-$,
\begin{equation}
\langle v_-|u_-\rangle=0\Rightarrow E_{\rm edge}=q\cos\theta,\;\;\text{if}\;\;-m<q\sin\theta<0.\label{edgemode}
\end{equation}
As illustrated in Fig.\ \ref{fig_plumbing}, the edge mode connects the bands of bulk modes in the TI, $E_{\rm TI}^2\geq q^2$, and in the MI, $E_{\rm MI}^2\geq q^2+m^2+2mq\sin\theta$. The connection is \textit{tangential}, so that the velocity $dE/dq$ changes continuously at the transition from an edge mode to a bulk mode \cite{tangentnote}.

The edge mode decays exponentially away from the boundary, with the same spinor structure on both TI and MI sides, but different decay lengths:
\begin{equation}
\psi(x)\propto \begin{pmatrix}
1-\sin\theta\\
i\cos\theta
\end{pmatrix}\times\begin{cases}
e^{-xq\sin\theta},&x<0,\\
e^{-x(m+q\sin\theta)},&x>0.
\end{cases}
\end{equation}
The decay length $\lambda_{\rm TI}=|q\sin\theta|^{-1}$ into the TI diverges when the edge mode merges with a bulk mode. Notice that there is no edge mode for purely normal magnetization: $\lambda_{\rm TI}\rightarrow\infty$ when $\theta\rightarrow 0$.	

\subsection{Infinite magnetization limit}

In the limit $m\rightarrow \infty$ of an infinite magnetization the decay length $\lambda_{\rm MI}=(m+q\sin\theta)^{-1}$ into the MI becomes vanishingly small. The eigenvalue problem can be restricted to the TI region $x<0$ with boundary condition
\begin{equation}
\psi(0)=\Omega\psi(0),\;\;\Omega=-m^{-1}\Xi_{\rm M}=\sigma_y\cos\theta-\sigma_z\sin\theta.\label{psiBC}
\end{equation}
The edge mode then extends over the half-infinite range $q\sin\theta<0$ --- the MI bulk bands are pushed to infinity.

Eq.\ \eqref{psiBC} is the boundary condition at a reconstructed zigzag edge (``reczag'' edge) of a graphene sheet \cite{Ost11}. In that context the edge mode is helical, it propagates in opposite directions in the two valleys \cite{Vol09,Tka09}. In the Brillouin zone of graphene the edge mode connects the Dirac points at opposite corners. In the TI there is only a single Dirac point and the edge mode has a single chirality.

\section{Dispersion relation in an MI--TI--MI channel}

A channel of width $W$, parallel to the $y$-axis, is created in an MI--TI--MI geometry, where the massless Dirac fermions are confined to the region $-W/2<x<W/2$. We first take the infinite magnetization limit, described by the boundary conditions
\begin{equation}
\begin{split}
&\psi(W/2,y)=\Omega(\theta_+)\psi(W/2,y),\\
&\psi(-W/2,y)=-\Omega(\theta_-)\psi(-W/2,y),\\
&\Omega(\theta)=\sigma_y\cos\theta-\sigma_z\sin\theta.
\end{split}
\label{channelBC}
\end{equation}
This corresponds to a boundary magnetization at an angle $\theta_\pm$ with the $z$-axis on the edge at $x=\pm W/2$. (The sign difference in the boundary condition at $x=\pm W/2$ appears because the outward normal $(\pm 1,0,0)$ changes sign.)

The wave function at the two boundaries is related by $\psi(W/2)=e^{\Xi_0 W}\psi(-W/2)$, with the matrix $\Xi_0$ given by Eq.\ \eqref{Xi0XiMdef} (at a given parallel momentum $q$). The Hermitian matrix $\Omega(\theta)$ has eigenvalue $\pm 1$ with eigenvector $e_\pm(\theta)=(-\sin\theta \pm 1,i\cos\theta)$. To satisfy the boundary condition we need $\psi(-W/2)$ parallel to $e_-(\theta_-)$ and $\psi(W/2)$ parallel to $e_+(\theta_+)$ hence orthogonal to $e_-(\theta_+)$:
\begin{equation}
\langle e_-(\theta_+)|e^{\Xi_0 W}|e_-(\theta_-)\rangle=0.\label{vpm}
\end{equation}
This works out as
\begin{subequations}
\label{Echannelq}
\begin{align}
& E\sin\alpha_- =\frac{k(E)\cos\alpha_-}{\tan[k(E)W]} + q\sin\alpha_+,\\
&k(E)=\sqrt{E^2-q^2},\;\;\alpha_\pm=\tfrac{1}{2}(\theta_-\pm\theta_+).
\end{align}
\end{subequations}

For $q=0$ the solution is
\begin{align}
\label{gapeq}
\tan EW&=\frac{1}{\tan\alpha_-}\\
&\Rightarrow\lim_{q\rightarrow 0}WE_n(q)=\pi/2-\alpha_-+n\pi,\;\;n\in\mathbb{Z}.\nonumber
\end{align}

\begin{figure}[tb]
\centerline{\includegraphics[width=1\linewidth]{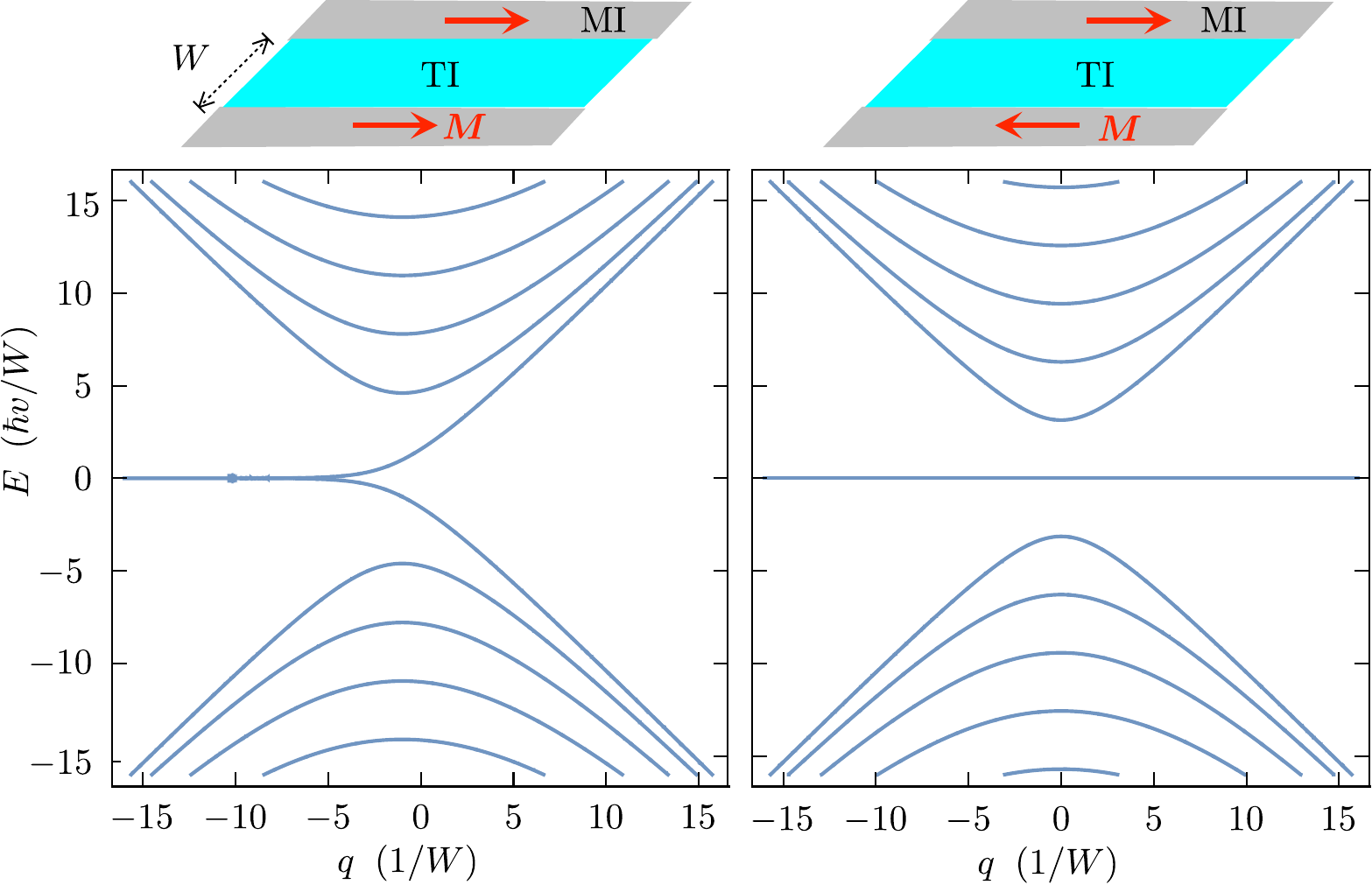}}
\caption{Band structure of an MI-TI--MI channel (width $W$), in the limit of infinitely large magnetization $\bm{M}$. The magnetization is oriented parallel to the MI--TI boundary. The left panel shows the case that $\bm{M}$ points in the same direction on opposite boundaries, in the right panel the direction on one boundary is inverted. The curves are calculated from Eq.\ \eqref{Echannelq}.
}
\label{fig_channel}
\end{figure}

\begin{figure*}[tb]
\centerline{\includegraphics[width=0.9\linewidth]{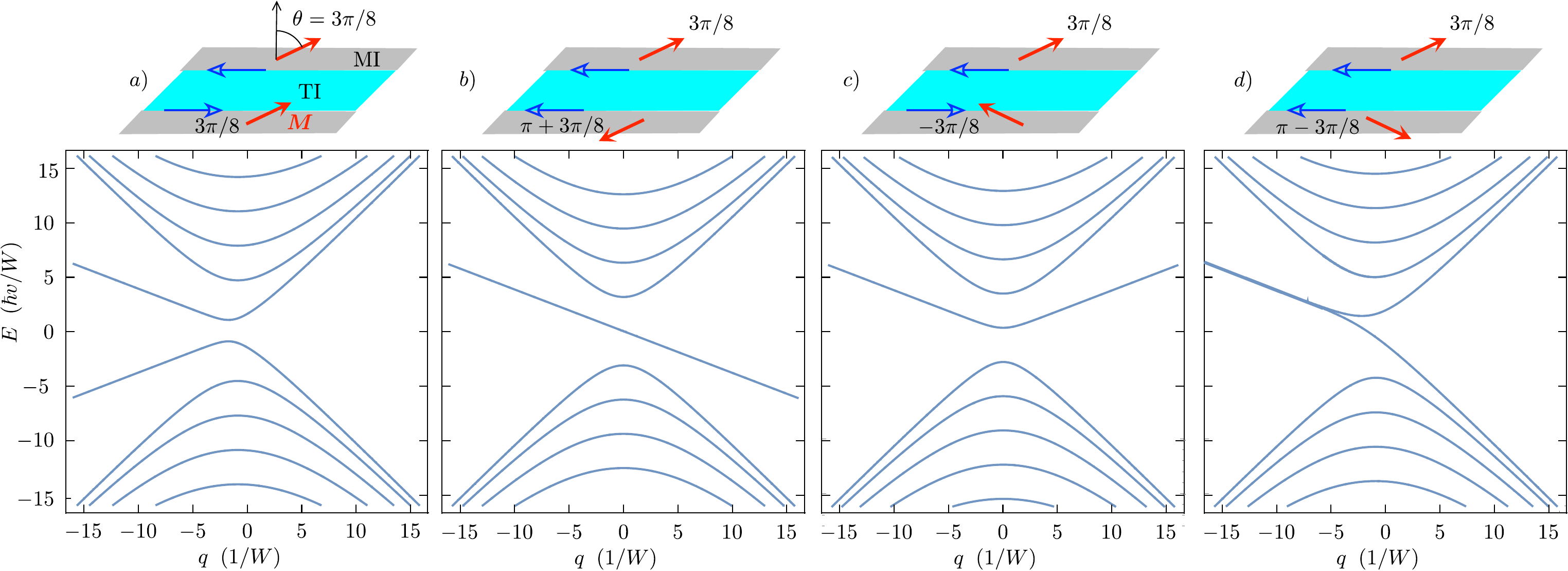}}
\caption{Same as Fig.\ \ref{fig_channel}, but now for a magnetization that has an out-of-plane component.
}
\label{fig_channel3}
\end{figure*}

Eq.\ \eqref{Echannelq} can be solved numerically for $E$ as a function of $q$. The spectrum consists of weakly $\theta$-dependent bulk modes, with bare velocity $v$ for large $|q|$, and a pair of edge modes $E_\pm$ with a reduced velocity \cite{note1} $v\cos\theta_\pm$,
\begin{equation}
E_\pm(q)=\pm q\cos\theta_\pm,\;\;\text{if}\;\;q\sin\theta_\pm <0.\label{Epmq}
\end{equation}

In Fig.\ \ref{fig_channel} we show the case $|\theta_\pm|=\pi/2$ of a magnetization that is parallel to boundary. The edge mode is then a dispersionless flat band, either connected to the bulk bands (if $\theta_+=\theta_-$) or disconnected (if $\theta_+=-\theta_-$). In graphene the corresponding band structure applies, respectively, to a nanoribbon with two zigzag edges or with one zigzag edge and one bearded edge \cite{Wak01,Koh07}. 

The flat bands acquire a dispersion if the magnetization has an out-of-plane component, see Fig.\ \ref{fig_channel3}. The gap in the spectrum near $q=0$ in panels a) and c) is a finite size effect, of order $\hbar v/W$ [see Eq.\ \eqref{gapeq}]. The gap vanishes if the normal component $M_z$ of the magnetization on opposite edges has the opposite sign [panels b) and d)].

If we relax the assumption of infinite magnetization, the wave function may penetrate into the magnetic insulator. We wrap the geometry on a cylinder along the $y$-axis, circumference $D+W$, so that the magnetic insulator extends over the two regions $W/2<x<(W+D)/2$ and $-(W+D)/2<x<-W/2$.   We take $\bm{M}=m(0,\sin\theta,\cos\theta)$ the same in both regions, and consider either periodic or anti-periodic boundary conditions: $\psi((W+D)/2)=\pm\psi(-(W+D)/2)$. 

\begin{figure}[tb]
\centerline{\includegraphics[width=0.8\linewidth]{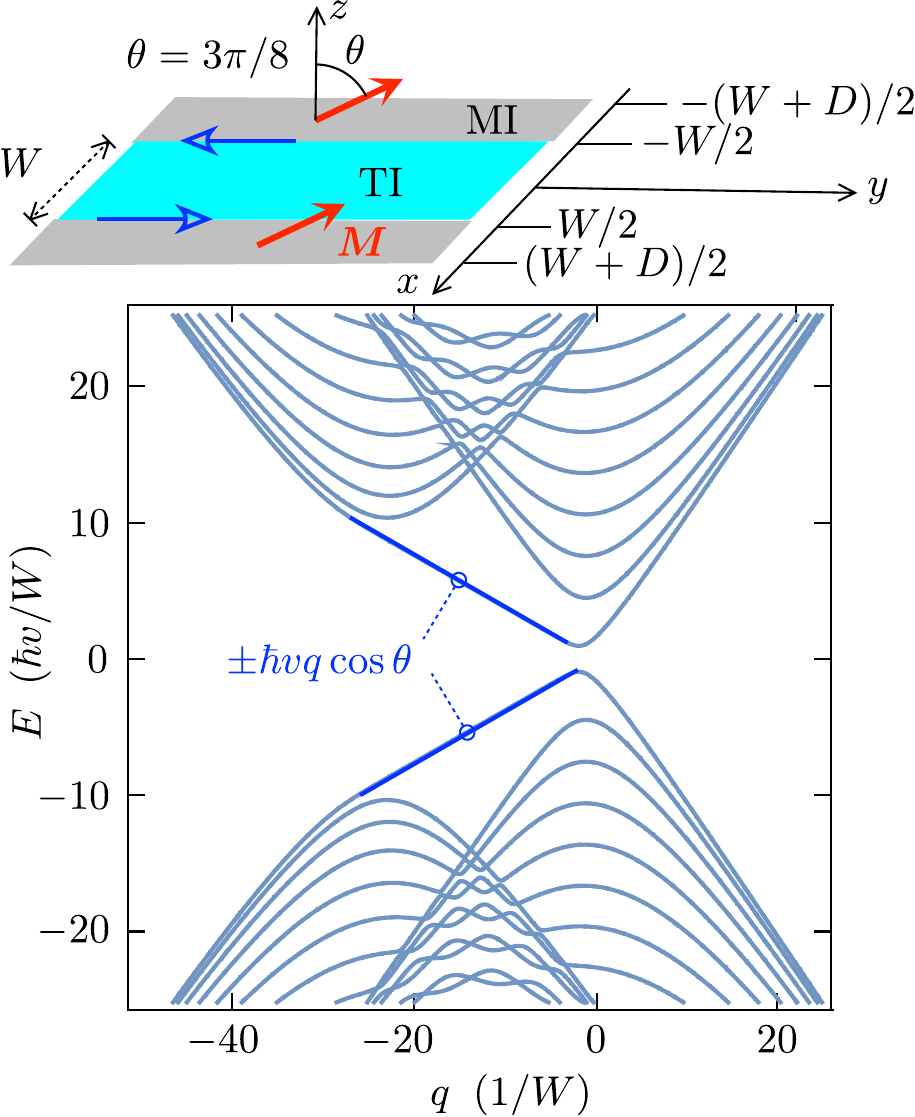}}
\caption{Same as Fig.\ \ref{fig_channel3}a, but now for a finite magnetization, at an angle $\theta=3\pi/8$ with the normal to the surface. The curves are calculated from Eq. \eqref{deteq}, for $W=D$, $m=25\,\hbar v/W$ (periodic boundary conditions at $x=\pm (W+D)/2$; anti-periodic boundary conditions give curves that are indistinguishable). The counterpropagating edge modes that connect bands in the TI and MI are indicated (dark blue).
}
\label{fig_channel2}
\end{figure}

Since
\begin{align}
 \psi((W+D)/2)={}&e^{(D/2)(\Xi_0+\Xi_{\rm M})}e^{W\Xi_0}e^{(D/2)(\Xi_0+\Xi_{\rm M})}\nonumber\\
 &\cdot\psi(-(W+D)/2),
\end{align}
the boundary condition implies the determinantal equation
\begin{subequations}
\begin{align}
&\det\bigl(1\mp e^{D(\Xi_0+\Xi_{\rm M})}e^{W\Xi_0}\bigr)=0\nonumber\\
&\Rightarrow \left(E^2-q^2-m q \sin \theta\right)\frac{ \sin  W k_{\rm TI} \sin D k_{\rm MI} }{k_{\rm TI} k_{\rm MI}}\nonumber\\
&\qquad=\cos  W k_{\rm TI} \cos D k_{\rm MI}\mp 1,\\
&k_{\rm TI}=\sqrt{E^2-q^2},\;\;k_{\rm MI}=\sqrt{E^2-q^2-m^2-2mq\sin\theta}.
\end{align}
\label{deteq}
\end{subequations}
(Check that Eq.\ \eqref{Echannelq} with $\alpha_-=0$, $\alpha_+=\theta$ is recovered in the $m\rightarrow\infty$ limit.)

In Fig.\ \ref{fig_channel2} we show how counterpropagating edge modes at opposite boundaries connect the TI and MI bands at positive and negative energies.

\section{Effect of a chemical potential mismatch}
\label{sec_deltamu}

\begin{figure}[tb]
\centerline{\includegraphics[width=0.6\linewidth]{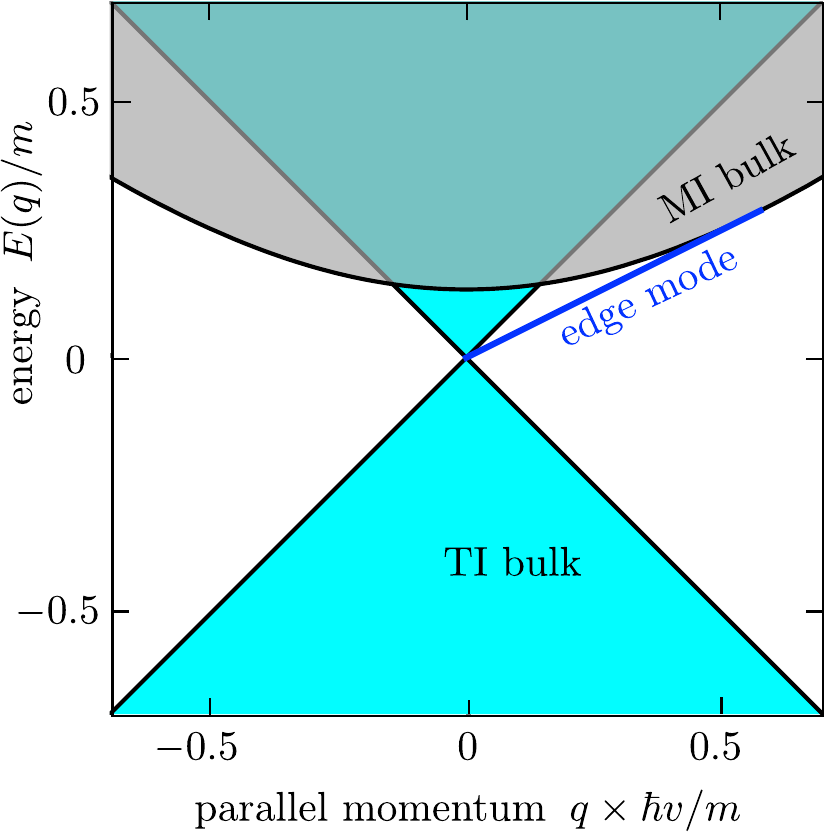}}
\caption{Same as Fig.\ \ref{fig_plumbing}, but now for $\theta=0$ at a chemical potential mismatch $\delta\mu=\tfrac{1}{2}m\sqrt{3}$, which enables the appearance of an edge mode at a purely perpendicular magnetization.
}
\label{fig_plumbing2}
\end{figure}

So far we have assumed that the chemical potential $\mu$ is the same on both sides of the TI--MI interface, $\delta\mu\equiv\mu_{\rm MI}-\mu_{\rm TI}=0$. Let us relax that assumption. The chemical potential  enters into Eq.\ \eqref{dpsidx1} as an energy-offset,
\begin{equation}
-i\sigma_x\psi'(x)=\begin{cases}
(E+\mu_{\rm TI}-q\sigma_y)\psi(x)&x<0,\\
\bigl(E+\mu_{\rm MI}-(q+m\sin\theta)\sigma_y\\
-m\sigma_z\cos\theta\bigr)\psi(x),&x>0.
\end{cases}\label{dpsidx2}
\end{equation}
The edge mode dispersion can then be calculated in the same way as in Sec.\ \ref{sec_edgemodedispersion}.

The edge mode persists if $|\delta\mu|<m$, in which case its effect is fully described by the angle $\phi=\arcsin(\delta\mu/m)$. Instead of Eq.\ \eqref{edgemode} we now have the dispersion relation
\begin{equation}
E_{\rm edge}=q\cos(\theta-\phi),\;\;\text{if}\;\;-m\cos\phi<q\sin(\theta-\phi)<0.\label{edgemode2}
\end{equation}
The inverse decay lengths into the TI  and MI are
\begin{equation}
\begin{split}
&\lambda_{\rm TI}^{-1}=|q\sin(\theta-\phi)|,\\
&\lambda_{\rm MI}^{-1}=m\cos\phi+q\sin(\theta-\phi).
\end{split}
\end{equation}
For $\theta=\phi$ mod $\pi$ the edge state merges with the bulk of the TI.

In Fig.\ \ref{fig_plumbing2} we show the edge mode for $\theta=0$, so for a fully perpendicular magnetization, in the presence of a chemical potential mismatch $\delta\mu=\tfrac{1}{2}m\sqrt{3}\Rightarrow\phi=\pi/3$. The arc state connecting the Dirac point to the magnetic band is clearly visible.

\section{Conclusion}

Confinement of 2D massless Dirac fermions by a mass boundary does not produce an edge mode \cite{Ber87}. This applies to the surface of a 3D topological insulator (TI) at an interface to a magnetic insulator (MI) with a perpendicular magnetization $M_\perp$. We have discussed two ways by which a chiral edge mode can be enabled: 
\begin{itemize}
\item by rotation of the magnetization $\bm{M}$ so that it has a component $M_\parallel$ parallel to the TI--MI interface;
\item by introduction of a chemical potential mismatch $\delta\mu<M$ across the interface.
\end{itemize}

The two mechanisms are described by a pair of angles $\theta=\arctan(M_\parallel/M_\perp)$ and $\phi=\arcsin(\delta\mu/M)$, which determine the edge mode velocity $\propto\cos(\theta-\phi)$ and the decay length $\propto 1/|\sin(\theta-\phi)|$ of the edge state into the TI.

The chiral edge mode can be observed in electrical conduction, similarly to the way the surface Fermi arc of a Weyl semimetal affects its transport properties \cite{Yan17}. One such effect is that it enables a Hall current density $j_x=(e^2/h)E_y$ into the gapped magnetic insulator, by means of the spectral flow induced by an electric field $E_y$ parallel to boundary. 

To see this, note that the electric field drives charge along the edge mode at a rate $dq/dt=(e/\hbar)E_y$. The edge mode extends over an interval $\Delta q$ and contains $\Delta q (L/2\pi)$ states, with $L$ the length of the system in the $y$-direction. After a time $\Delta t=(dq/dt)^{-1}\Delta q$ a charge $\Delta Q=e\Delta q(L/2\pi)$ has been transferred between MI and TI, producing a current density
\begin{equation}
j_x=\frac{1}{L}\frac{\Delta Q}{\Delta t}=\frac{e}{2\pi}\frac{dq}{dt}=\frac{e^2}{h}E_y.\label{jxresult}
\end{equation}
This effect is related to the anomalous quantum Hall effect of a topological insulator, studied extensively in theory \cite{Wat10,Chu11,Koe14,Zho22,Zou22,Bai23}, in electrical conduction experiments \cite{Xu14,Mog22}, and in Faraday rotation optical experiments \cite{Wu16,Oka16}.

\acknowledgments

This project has received funding from the European Research Council (ERC) under the European Union's Horizon 2020 research and innovation programme. Discussions with A. R. Akhmerov, F. Hassler, and J. E. Moore are gratefully acknowledged.


\begin{thebibliography}{99}
\bibitem{note1} The edge mode dispersion \eqref{Epmq} follows from Eq.\ \eqref{Echannelq} for imaginary $k=i\kappa$, $\kappa W\gg 1$, so that $k/\tan kW\rightarrow \kappa$.
\end{thebibliography}
\end{document}